%% file: cameraready_v2.tex
\documentclass{article}
\usepackage{spconf,amsmath,graphicx}
\usepackage{amsthm}
\usepackage{thmtools, thm-restate}
\usepackage{algorithm}
\input{symbols_commands}
\usepackage{cite}
\usepackage{amssymb,amsfonts}
\usepackage{algorithmic}
\usepackage{graphicx}
\usepackage{textcomp}
\usepackage{xcolor}
\usepackage{units}
\usepackage{multirow}
\usepackage{colortbl}
\usepackage{soul}

\usepackage[T1]{fontenc}
\usepackage[utf8]{inputenc}

\usepackage{caption}
\usepackage[font={small}]{caption}

\allowdisplaybreaks[4]

\AtBeginDocument{
    \setlength{\abovedisplayskip}{1pt}
    \setlength{\belowdisplayskip}{1pt}
    \setlength{\abovedisplayshortskip}{0pt}
    \setlength{\belowdisplayshortskip}{0pt}
}

\newcommand{\alg}{
$\mathsf{ROAR-Fed}~$}

\newcommand{\algns}{$\mathsf{ROAR-Fed}$}
\newcommand{\algp}{$\mathsf{PROAR-PFed}$}
\newcommand{\alghota}{$\mathsf{HOTA-FedGradNorm}$}

\title{Personalized Over-the-Air Federated Learning with Personalized Reconfigurable Intelligent Surfaces}
\name{Jiayu Mao and Aylin Yener\thanks{This work is supported in part by NSF CNS-2112471.}}
\address{Dept. of Electrical and Computer Engineering, The Ohio State University}
\begin{document}
\maketitle
\begin{abstract}
Over-the-air federated learning (OTA-FL) provides bandwidth-efficient learning by leveraging the inherent superposition property of wireless channels. Personalized federated learning balances performance for users with diverse datasets, addressing real-life data heterogeneity. We propose the first personalized OTA-FL scheme through multi-task learning, assisted by personal reconfigurable intelligent surfaces (RIS) for each user. We take a cross-layer approach that optimizes communication and computation resources for global and personalized tasks in time-varying channels with imperfect channel state information, using multi-task learning for non-i.i.d data. Our PROAR-PFed algorithm adaptively designs power, local iterations, and RIS configurations. We present convergence analysis for non-convex objectives and demonstrate that PROAR-PFed outperforms state-of-the-art on the Fashion-MNIST dataset.
\end{abstract}
\vspace{-0.2in}
\section{Introduction}
\label{sec:intro}
\vspace{-0.12in}

Federated learning (FL) \cite{mcmahan2017} is a popular distributed machine learning paradigm that employs collaborative iterative training between a parameter server (PS) and edge devices (clients). During each iteration, clients train local models, which are subsequently sent to the PS for aggregation and global model update. Whilst a fitting paradigm for mobile edge networks, addressing challenges brought on by the radio channel and limited wireless resources is essential for FL.
Over-the-air federated learning (OTA-FL)~\cite{amiri2020} leverages the broadcast nature of the wireless channel for bandwidth-efficient learning, by having all clients simultaneously transmit analog model updates, enabling the PS to directly receive the aggregated model through the intrinsic superposition over the air.
Naturally, over-the-air (OTA) aggregation relies on transmitter-side channel state information (CSI) for power control.
In practice, users only have access to estimated CSI, potentially harming learning performance. This work quantifies the impact of estimated CSI on performance.

Personalized federated learning (PFL) is a recent framework designed to tackle FL scenarios where non-i.i.d. datasets can cause severe performance degradation for individual users. Despite having attracted substantial attention in machine learning \cite{fallah2020personalized,li2021ditto,xue22diple}, PFL approaches are rare in wireless systems: reference  \cite{sami2022over} utilizes clustered FL; \cite{mortaheb2022personalized},~\cite{chen2023personalizing} explore multi-task learning over-the-air in a hierarchical setup.
By contrast, we propose a genuine PFL algorithm for each individual client integrating the wireless physical layer with OTA computation. 

Reconfigurable intelligent surfaces (RIS)~\cite{wu2019intel} are programmable meta-surfaces with low-cost passive reflecting elements that adjust phase shifts of incident signals.
When integrated with OTA-FL, RIS can create more favorable propagation environments and facilitate better model aggregation. 
Existing research~\cite{ni2021fed,wang2021fed,li2022one} mainly focuses on minimizing the mean squared error (MSE) of FL model aggregation, while~\cite{liu2021risfl} concentrates on unified communication and learning design based on a time-invariant static channel with perfect CSI.
Recently, we have developed an adaptive joint communication and learning algorithm in time-varying channels assisted by one RIS~\cite{mao22roar,mao22iccw}.
All of these works consider solving a single global FL problem.

In this paper, we bring personalization into OTA-FL and 6G programmable wireless environments. We propose the \emph{first} personalized OTA-FL framework with assistance from \emph{personal} RIS for each client.
Envisioning 6G with portable and compact RIS units on intelligent edge devices, we explore and validate the personal RIS model's potential for personalized learning objectives, equipping users with individual RIS under time-varying physical layers and imperfect CSI.
Inspired from our prior works~\cite{mao22roar,mao22iccw} for a single global learning objective and single RIS, we establish an alternating, cross-layer approach optimizing resources for enhanced global and personalized learning, in a system model that personalizes the physical layer.
Different than existing personalized OTA-FL works that assume strong convexity, our framework is convergent under non-convex objectives and device heterogeneity.
Specifically, we propose \algp~(\underline{P}ersonal \underline{R}IS-assisted \underline{O}ver-the-\underline{A}ir \underline{R}esource Allocation for \underline{P}ersonalized \underline{Fed}erated Learning), which adaptively designs power control, local training iterations and {\it personal} RIS configurations during each global iteration, to enable each client's {\it personalized} model training together with global learning objective at no additional cost.
We present the convergence analysis of~\algp~and assess its performance on the Fashion-MNIST dataset with imperfect CSI, showing its superior performance and effectiveness for personalization.

\vspace{-0.2in}
\section{System Model} 
\label{sec: prelim}
\vspace{-0.15in}

\begin{figure}[t] 
    \centering
    \includegraphics[scale=0.25]{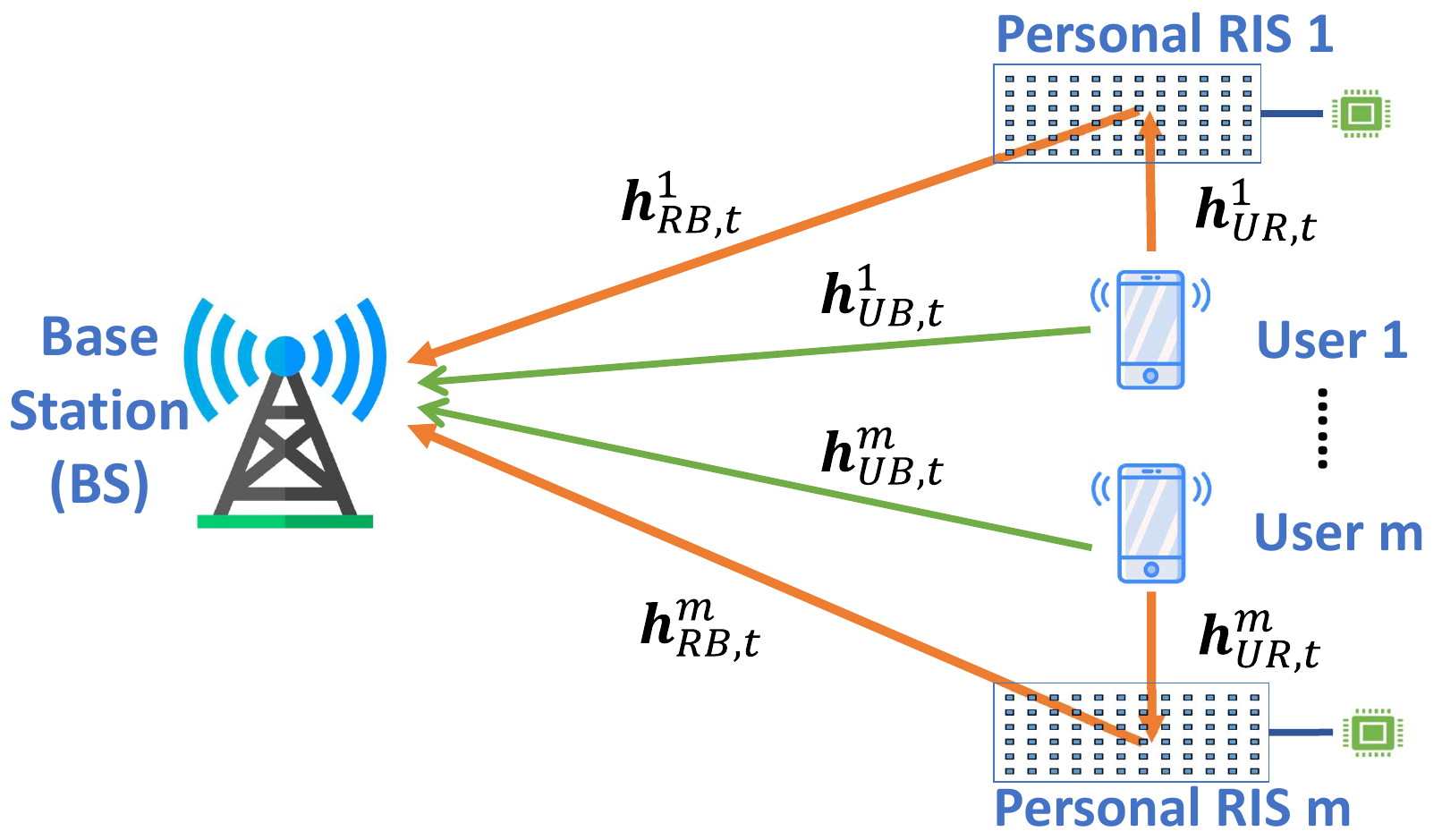}
    \vspace{-0.1in}
    \caption{The personal RIS-assisted communication system.}
    \label{fig:sysmodel}
    \vspace{-0.1in}
\end{figure}

We consider a personal RIS-assisted communication system with $m$ clients and single antenna PS (Fig.~\ref{fig:sysmodel}).
Each client $i$ has a training dataset $D_i$ with distinct distribution $\mc{X}_i$, i.e., $\mc{X}_i \neq \mc{X}_j$ if $ i \neq j, \forall i, j \in [m]$.
The goal of conventional FL is to solve a single global objective:
\begin{equation}
    \min_{\w \in \mathbb{R}^d}F(\w) \triangleq \min_{\w\in\mathbb{R}^d} \sum_{i \in [m]} \alpha_i F_i(\w, D_i), 
    \label{equ: objective}
\end{equation}
where $\w \in \Rb^d$ denote the model, $F_i(\w, D_i)$ is local objective function, $\alpha_i = \frac{| D_i |}{\sum_{i \in [m]} | D_i |}$ is the weight of client $i$.

To better accommodate data heterogeneity, in this paper, we consider a bi-level personalized federated learning through a multi-task learning framework:
\begin{equation}
    \begin{aligned} 
        & \min_{\mathbf{v}^i \in \mathbb{R}^d}R_i(\mathbf{v}^i;\w^*) \triangleq F_i(\mathbf{v}^i, D_i) + \frac{\lambda}{2} \|\mathbf{v}^i - \w^*\|^2  \\
        &\begin{array}{ll}
        s.t. & \w^* = \underset{\w}{\mathrm{arg\,min}} F(\w),
        \end{array}
    \end{aligned}
    \label{equ: localobjective}
\end{equation}
where $\mathbf{v}^i$ is the local parameter of user $i$ for personalized task, and $\lambda$ is the hyperparameter of regularization. When $\lambda \rightarrow 0$, the personalized task reduces to local model training; while $\lambda \rightarrow \infty$, it becomes conventional FL.
In FL, clients minimize $F_i$ and send the local updates to the PS.
OTA-FL enables model aggregation via concurrent transmissions over the wireless medium. Next, PS broadcasts the updated global model to all devices. This procedure continues until convergence. In particular, for local training, client $i$ performs stochastic gradient descent (SGD) to calculate its local gradient with the global model initialization $\w_t$ and its dataset $D_i$ for $\tau_t^i$ steps, which varies across different rounds and users, consistent with our earlier works~\cite{mao22roar,mao22iccw}.

We consider weak direct links between devices and PS, rendering RIS assistance essential.
We assume that each device uses an individual RIS for uplink, with negligible interference from other RISs.
Assuming uniform $N$ elements for each RIS, we adjust their phase shifts per global iteration. 
The phase matrix of RIS $i$ in round $t$ is $\The_t^i = diag (\theta_{1,t}^i, \cdots, \theta_{N,t}^i)$, with $\theta_{n,t}^i = e^{j \phi_{n,t}^i}$.

We assume a block fading model for uplink, channel coefficients stay unchanged within a communication round, but vary independently between rounds.
We consider an error-free downlink for simplicity, where clients receive an accurate global model per iteration, i.e., $\w^i_{t, 0} = \w_t, \forall i\in [m]$. We note that our results readily extend to noisy downlinks by including dynamic power control for downlinks as done in~\cite{mao22iccw}.
Let $\h_{RB,t}^i \in \Cb^N$, $\h_{UR,t}^{i} \in \Cb^N$, $h_{UB,t}^{i} \in \Cb$ be the channel gains from RIS $i$ to PS, from user $i$ to RIS $i$, and from user $i$ to PS, respectively.
Denote $\x_t^i \in \mb{R}^d$ as transmit signal of user $i$, $\z_t$ as the additive white Gaussian noise (AWGN) with zero mean and variance $\sigma_c^2$.
The received signal $\y_t$ at the PS is:
\begin{equation}
    \y_t = \sum_{i \in [m]} (h_{UB,t}^{i} + (\h_{UR,t}^{i})^H \The_t^i \h_{RB,t}^i)\x^i_t + \z_t.\label{equ:receivesig}
\end{equation}
We further define the cascaded device $i$-RIS $i$-PS channel as $\g_t^i$, i.e., $\g_t^i = ((\h_{UR,t}^{i})^H \Hb_{RB,t}^i)^H \in \Cb^N$, where $\Hb_{RB,t}^i = diag(\h_{RB,t}^i)$.
Then, we can rewrite the phase matrix as a vector $\theb_t^i=(\theta_{1,t}^i,...,\theta_{N,t}^i)^T$.
Thus, the equivalent received signal is $\y_t = \sum_{i \in [m]} (h_{UB,t}^{i} +  (\g_t^i)^H \theb_t^i)\x^i_t + \z_t.$
The power constraint of device $i$ in round $t$ is $\mathbb{E}[\| \x^i_t \|^2] \leq P_t^i, \forall i \in [m], \forall t,$ where $P_t^i$ is the transmit power budget.

We have only imperfect CSI available at clients.
$\hh_t$ is the CSI estimate of each wireless path in $t$-th round: $\hh_t = h_t + \Delta_t, \forall t,$
where $\Delta_t$ represents the i.i.d. channel estimation error, with zero mean and variance $\sigt_h^2$.
Note that all links have this estimation error.
To simplify notation, we denote the overall channel gain of device $i$ in the $t$-th round as $h_t^i$:
\begin{equation}
    h_t^i=h_{UB,t}^{i} +  (\h_{UR,t}^{i})^H \The_t^i \h_{RB,t}^i.
\end{equation}
Similarly, $\hh_t^i$ is the overall estimated CSI at the device $i$.

\vspace{-0.2in}
\section{Algorithm Design} 
\label{sec: alg}
\vspace{-0.15in}

In this section, we detail Algorithm~\ref{alg:apaf} for joint communication and learning optimization. Each global iteration consists of three phases: personal RIS phase design, global model updates, and personalized model adjustments.

\begin{algorithm}[t!] 
    \caption{Personal RIS-assisted Over-the-Air Resource Allocation for Personalized Federated Learning~(\algp)} \label{alg:apaf} 
    \begin{algorithmic}[1]
    \STATE 
    \emph{\bf Initialization: $\w_0$, $\theb_0^i$, $\mathbf{v}_0^i$, $\beta_t^i$, $\tau_v^i$, $\tau_t^i,\forall i \in [m]$.}
    \FOR{$t=0, \dots, T-1$}
        \FOR{each device $i \in [m]$ in parallel}
        \FOR{$j=0, \dots, J-1$}
        \STATE {Each device updates its personal RIS $i$ by~\eqref{equ:phase}}.
        \ENDFOR
        \ENDFOR
        \STATE {PS broadcasts the global model $\w_t$.}
        \FOR{each device $i \in [m]$ in parallel}
        \STATE {Each device gets $\tau_t^i$ to satisfy the power constraint and starts local training, finds $\beta_t^i$ and transmits $\x_t^i$.}
        \STATE {Each device updates $\mathbf{v}_t^i$ for $\tau_v^i$ local steps.}    
        \ENDFOR
    \STATE {The PS aggregates and updates global model by~\eqref{equ:globalup}.}
    \ENDFOR
    \STATE {\textbf{return} $\{\mathbf{v}^i\}_{i \in [m]} (personalized), \w_T (global)$}
    \end{algorithmic}
\end{algorithm}

We employ dynamic power control (PC) for the global model update~\cite{yang22,mao22,mao22roar,mao22iccw}.
Let $\beta_t^i$ and $\beta_t$ be the PC factor for device $i$ and PS, respectively. 
Device $i$ gets the local updates and computes signal $\x_t^i$ to transmit:
\begin{equation}
    \x_t^i = \beta_t^i (\w^i_{t, \tau_t^i} - \w^i_{t, 0}), \forall t. \label{equ:transig}
\end{equation}
Next, the PS applies $\beta_t$ to the received signal~\eqref{equ:receivesig}: 
\begin{align}
    \w_{t+1} &= \w_{t} + \frac{1}{\beta_t}\sum_{i=1}^{m} h_t^i \x_t^i  + \tilde{\z}_t, \tilde{\z}_t \sim \mc{N}(\0, \frac{\sigma_c^2}{\beta_t^2} \mf{I}_d). \label{equ:globalup}
\end{align}
We propose a channel inversion strategy combined with dynamic local steps to alleviate the effects of channel fading.
Specifically, we design two criteria for each edge device $i$:
\begin{equation} \label{equ:betaiimp}
    \beta_t^i = \frac{\beta_t \alpha_i}{\tau^i_t \hh_t^i}, \quad 3\eta_t^2 \beta_t^i \tau_t^i G^2 \leq P_t^i,
\end{equation}
where $G$ is the bound of the stochastic gradient.
Using dynamic local steps, $\tau_t^i$, we counter learning degradation from imperfect CSI-induced misalignment and leverage local computation resources. 
Criterion~\eqref{equ:betaiimp} is set to design phase shifts of RIS $i$, ensuring convergence.
After phase updates, each device $i$ finds $\tau_t^i$ by incorporating~\eqref{equ:betaiimp} and~\eqref{equ:transig} into the power constraint. Then it starts local training.

We assume that personal RISs are controlled by their users only.
Unlike our previous studies with a single RIS requiring user selection for phase updates, this work lets each device $i$ directly update its RIS according to criterion~\eqref{equ:betaiimp}.
The first step is to rewrite~\eqref{equ:betaiimp}:
\begin{equation} \label{inequ:phase}
    (\g_t^i)^H \theb_t^i \geq \frac{3 \eta_t^2 \beta_t \alpha_i G^2}{P_t^i} - \hh_{UB,t}^i.
\end{equation}
Then we design the phase as follows:
\begin{equation} \label{prob:phase}
    \begin{aligned} 
        & \mathop{min}\limits_{\theb_t^i} \quad \|(\g_t^i)^H \theb_t^i - \frac{3 \eta_t^2 \beta_t \alpha_i G^2}{P_t^i} + \hh_{UB,t}^i \|_2^2  \\
        &\begin{array}{ll}
        s.t. & |\theta_{t,n}^i|=1 , \quad n=1,...,N.
        \end{array}
    \end{aligned}
\end{equation}
(\ref{prob:phase}) is non-convex and we use successive convex approximation (SCA)~\cite{scutari2013,mao22papa}.
First, we define:
\begin{equation} \label{equ:phasefunc}
\begin{array}{ll}
     f(\theb_t^i) & = || s_t^i - (\g_t^i)^H \theb_t^i||_2^2   \\
     & = (s_t^i)^* s_t^i - 2 Re \{ (\theb_t^i)^H \textbf{a}\} + (\theb_t^i)^H \textbf{U} \theb_t^i ,
\end{array}
\end{equation}
where $\textbf{a} = s_t^i \g_t^i$, $\textbf{U} = \g_t^i (\g_t^i)^H$, $s_t^i = \frac{3 \eta_t^2 \beta_t \alpha_i G^2}{P_t^i} - \hh_{UB,t}^i$.
Using the equivalent phase element expression $\theta_{n,t}^i = e^{j \phi_{n,t}^i}$, and noting that $s_t^i$ is constant, we derive:
\begin{equation}
    f_1(\phib_t^i) = (e^{j\phib_t^i})^H \textbf{U} e^{j\phib_t^i} - 2 Re\{(e^{j\phib_t^i})^H \textbf{a}\},
\end{equation}
where $\phib_t^i = (\phi_{1,t}^i,...,\phi_{N,t}^i)^T$.
Next, we apply the SCA and use the second-order Taylor expansion to find the surrogate function $g(\phib_t^i,\phib_{t,j}^i)$ at point $\phib_{t,j}^i$ in iteration $j$, then use SGD to find the stationary solution $\phib_{t,J}^{i}$:
\begin{equation}
    \phib_{t,j+1}^{i} = \phib_{t,j}^{i} - \frac{\nabla f_1(\phib_{t,j}^{i})}{\lambda}. \label{equ:phase}
\end{equation}
Finally, we get the phase design of RIS $i$: $\theb_t^i= e^{j\phib_t^i}$.

Next, each device $i$ optimizes the personalized task in~\eqref{equ: localobjective}.
Instead of directly finding $\w^*$ to minimize $R_i(\mathbf{v}^i;\w^*)$, we adopt an alternating approach inspired by~\cite{li2021ditto} to solve the local objective approximately in each global round (see~\algp).
Specifically, device $i$ performs $\tau_v^i$ SGD steps, initializing with $\mathbf{v}_t^i$ from last global iteration:
\begin{equation}
    \mathbf{v}^i_{t, k+1} = \mathbf{v}^i_{t, k} - \eta_v (\nabla F_i(\mathbf{v}^i_{t, k}) + \lambda(\mathbf{v}^i_{t, k} - \w_t)), \forall k,\label{equ:persgd}
\end{equation}
where $\eta_v$ is the local learning rate, and we set $\mathbf{v}^i_{t+1}=\mathbf{v}^i_{t, \tau_v^i}$.
In round $t$, we use $\w_t$ to approximate $\w^*$ and each device updates independently. Thus, we can schedule this stage after transmission, letting devices use PS aggregation downtime for personalized training, saving overall learning time.


\vspace{-0.1in}
\section{Convergence Analysis}
\label{sec: conv}
\vspace{-0.1in}

We first provide the assumptions on non-convex objectives:
\vspace{-0.05in}
\begin{assum}\label{a_smooth}
	$\exists L > 0$, $ \| \nabla F_i(\w_1) - \nabla F_i(\w_2) \| \leq L \| \w_1 - \w_2 \|$, $\forall \w_1, \w_2 \in \mathbb{R}^d$, $\forall i \in [m]$.
\end{assum}
\vspace{-0.1in}
\begin{assum}\label{a_unbias}
	Local stochastic gradients are unbiased with bounded variance, i.e.,
	$\mathbb{E} [\nabla F_i(\w, \xi_i)] = \nabla F_i(\w)$, $\forall i \in [m]$, and $\mathbb{E} [\| \nabla F_i(\w, \xi_i) -  \nabla F_i(\w) \|^2] \leq \sigma^2$, where $\xi_i$ is sampled from $D_i$. Also, $\mathbb{E} [g_i(\mathbf{v}^i;\w)] = \nabla R_i(\mathbf{v}^i;\w)$, where $g_i(\mathbf{v}^i;\w)$ is stochastic gradient of $R_i(\mathbf{v}^i;\w)$.
\end{assum}
\vspace{-0.1in}
\begin{assum}\label{a_bounded}
    $\exists G \geq 0$, $\mathbb{E} [\| \nabla F_i(\w, \xi_i) \|^2] \leq G^2$, $\forall i \in [m]$.
\end{assum}
\vspace{-0.05in}
\textbf{Convergence of Algorithm~\ref{alg:apaf}.} The global model $\w$ does not rely on any personalized models $\{\mathbf{v}^i\}_{i \in [m]}$. Thus, the global optimization has the same convergence rate with \alg in our previous work~\cite{mao22roar}\footnote{When we extend to noisy downlink, the convergence of the global model aligns with~\cite{mao22iccw} without impacting personalized task convergence.}.

Based on this observation, we can now present the convergence analysis of the personalized local task as follows:
\vspace{-0.2in}
\begin{restatable} {theorem} {convergence} \label{thm:convergence}
    With Assumptions~\ref{a_smooth}-~\ref{a_bounded}, a constant global learning rate $\eta_t = \eta  \leq \frac{1}{L}$, a constant local learning rate $\eta_v \leq \frac{1}{\sqrt{2L^2 + 2 \lambda^2}}$ and $T\geq 4$, for each device $i\in [m]$, we have:
    \begin{multline}
        \min_{t \in [T]} \mb{E} \| \nabla R_i(\mathbf{v}_t^i;\w^{*}) \|^2 \leq 
        \underbrace{\sqrt{2L^2 + 2 \lambda^2} \eta_v^2 \sigma^2}_{\mathrm{statistical \, error}}  \nonumber \\
        +\underbrace{ \frac{1}{T} \sum_{s=0}^{T-1}\frac{2 \left(R_i(\mathbf{v}_s^i;\w_s) - R_i(\mathbf{v}_{s+1}^{i};\w_s) \right)}{\tau_v(\eta_v - \frac{\sqrt{2L^2 + 2 \lambda^2}}{2}\eta_v^2)} }_{\mathrm{optimization \, error}}   \underbrace{+ \lambda^2 T^2 \frac{\sigma_{c}^2}{\beta^2}}_{\mathrm{channel \, noise \, error}} \nonumber \\
        + \underbrace{2 \lambda^2 m G^2 \eta^2\frac{1}{T} \sum_{s=0}^{T-1}  (T-s) \sum_{l=s}^T \sum_{i=1}^m \alpha_i^2 \mb{E}_t \bigg\| \frac{h_l^i}{\hh_l^i} \bigg\|^2}_{\mathrm{global \, model \, update \, error}} \nonumber
    \end{multline}
    where $\frac{1}{\bar{\beta}^2} = \frac{1}{T} \sum_{t=0}^{T-1} \frac{1}{\beta_t^2}$, $\tau_v = \tau_v^i$.
\end{restatable}
\vspace{-0.15in}

\begin{proof}[Proof Highlights] 
We use Cauchy-Schwartz inequality to prove that the personalized objective $R_i$ is $\sqrt{2L^2 + 2\lambda^2}$-Lipschitz continuous with a variance $\sigma^2$ of stochastic gradients.
Through the local update rule and properties of the global update, channel noise, and gradients, we get the convergence bound after averaging over iterations.
\end{proof}
\vspace{-0.1in}

Theorem~\ref{thm:convergence} highlights four errors affecting convergence: statistical error, local optimization error, uplink channel noise, and global model update error. 
While the first two are common in non-convex cases, the latter two are coupled with global training because of our alternating scheme, making uplink imperfect CSI and noise influential in local optimization.

The convergence upper bound is finite:
as the channel estimation error is typically small, we adopt the Taylor expansion similar to \cite{zhu2020one}.
By ignoring higher-order terms and selecting the proper hyperparameters, we obtain:
\vspace{-0.05in}
\begin{restatable}{corollary} {convergence_rate} \label{cor:convergence}
Let $|\Delta_t| \ll |h_t|, \forall t\in[T]$, $h_{UB,m} = \\
\mathop{min}\limits_{t \in [T], i \in[m]}\{|h_{UB,t}^i|\}$, $h_{UR,a} = \mathop{max}\limits_{t \in [T], i \in[m], j \in [N]}\{|h_{UB,t,j}^i|\}$, $h_{RB,a} = \mathop{max}\limits_{t \in [T], j \in [N]}\{|h_{RB,t,j}|\}$, $\eta = \frac{1}{T}$, $\beta = T$, $\tau_v^i = \tau_v$, $\alpha_i=\frac{1}{m}$, $\exists \lambda < \epsilon, \epsilon>0,$ the convergence rate is bounded:
\begin{align}
    &\min_{t \in [T]} \mb{E} \| \nabla R_i(\mathbf{v}_t^i;\w^{*}) \|^2 \leq \sqrt{2L^2 + 2\lambda^2} \eta_v^2 \sigma^2 + \lambda^2 \sigma_{c}^2+\nonumber \\
    &\frac{1}{T} \sum_{s=0}^{T-1}\frac{2 \left(R_i(\mathbf{v}_s^i;\w_s) - R_i(\mathbf{v}_{s+1}^{i};\w_s) \right)}{\tau_v(\eta_v - \frac{\sqrt{2L^2 + 2 \lambda^2}}{2}\eta_v^2)}  + 2 \lambda^2 G^2(1+C) \nonumber,
\end{align}
where $C=\frac{\sigt_h^2 (1+N^2(h_{UR,a}^2+h_{RB,a}^2+\sigt_h^2))}{(h_{UB,m})^2}$. 
\end{restatable}
\vspace{-0.05in}
The number of personal RIS elements $N$ has an impact on the convergence bound of personalized tasks.
While a larger $N$ can increase channel estimation errors, it does not dominate the bound as it is tied to small magnitudes of $\lambda$ and $\sigt_h^2$.

\vspace{-0.2in}
\section{Numerical Results} 
\label{sec: exp}
\vspace{-0.15in}
\begin{figure}[t] 
    \centering
    \includegraphics[scale=0.48]{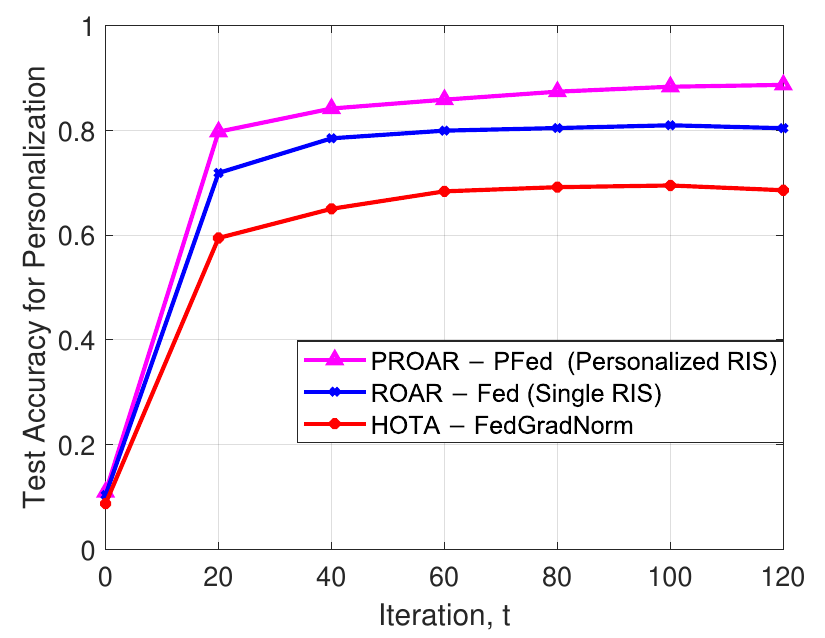}
    \vspace{-0.1in}
    \caption{Average test accuracy for personalized tasks when $\gamma=0.5$.}
    \label{fig:result1}
    \vspace{-0.1in}
\end{figure}
We simulate a 3D multiuser communication system with personalized RISs.
We have $m=10$ clients, each with a RIS of $N=10$ elements.
Similar to~\cite{liu2021risfl}, we place PS at $(-50,0,10)$ and users are uniformly spread in the x-y plane, with $x\in[-20,0]$ and $y \in [-30,30]$.
RISs are located two meters above their respective users. 
We consider i.i.d. Gaussian fading channels and a path loss model from~\cite{tang2020ris}.
The path loss for the direct link is $ G_{PS}G_{U}\left(\frac{3*10^8 m/s}{4 \pi f_c d_{UP}} \right)^{PL}$, where $G_{PS}=5$dBi, $G_U=0$dBi, $f_c=915$MHz, the user-PS distance is $d_{UP}$, and the path loss exponent $PL=4$.
For RIS-assisted link, it is $G_{PS}G_{U} G_{RIS} \frac{N^2 d_x d_y ((3*10^8 m/s)/f_c)^2 }{64 \pi^3 d_{RP}^2 d_{UR}^2}$, where $G_{RIS}=5$dBi, $d_x=d_y=(3*10^7 m/s)/f_c$, RIS-PS distance is $d_{RP}$, and user-RIS distance is $d_{UR}$. 
We model the channel estimation error as Gaussian with $\sigt_h^2= 0.1 \sigma_c^2$, and transmit SNR is 20dB.
We perform image classification on the Fashion-MNIST dataset~\cite{xiao2017fashion} with a CNN featuring two $5 \times 5$ convolution layers (10 and 20 channels), batch normalization, a 50-unit fully connected layer, and a softmax output.
Parameters are set with $\lambda=0.1$ and $\tau_v^i = 3$.
We use non-i.i.d. data partitioning with a Dirichlet distribution $Dir_{10}(\gamma)$ as in~\cite{li2022federated}.
We benchmark: 1) \algns~\cite{mao22roar}, employing a central RIS at $(0,0,10)$ meters with $m \times N$ elements~\cite{ni2021fed} for global model; 2) \alghota~\cite{mortaheb2022personalized}, a hierarchical OTA-FL without RIS. The system features PS, intermediate servers, and clients in $m$ clusters with fading PS-server links and error-free server-client connections.
Each client's model has shared global layers and an individual layer, focusing on personalization.

\begin{table}[t]
\caption{Average (5 trails) of the CNN test accuracy (\%) comparison on Fashion-MNIST with various non-i.i.d. level $\gamma$. We report \textbf{average} across users for personalized models. “N/A” means not applicable.}
\label{table:test_accuracy}
\centering
{\scriptsize
\begin{tabular}{|c|c|cc|}
\hline
\multicolumn{1}{|c|}{\multirow{2}{*}{\textbf{Non-IID}}} & \multicolumn{1}{c|}{\multirow{2}{*}{\textbf{Algorithm}}} & \multicolumn{2}{c|}{\textbf{FL Model}}                                        \\ \cline{3-4} 
\multicolumn{1}{|c|}{}                               & \multicolumn{1}{c|}{}                           & \multicolumn{1}{c|}{Global}  & \multicolumn{1}{c|}{Personalized}    \\ \hline
\multirow{3}{*}{$\gamma = 0.1$}                                 & {\cellcolor[gray]{.9}{\textbf{\algp} }}                                     & \multicolumn{1}{c|}{\cellcolor[gray]{.9} {\bf 63.08}} & \multicolumn{1}{c|}{\cellcolor[gray]{.9} {\bf 96.04 $\pm$ 2.69}}  \\ \cline{2-4} 
                                                     & \alg                                           & \multicolumn{1}{c|}{57.68} & \multicolumn{1}{c|}{88.97 $\pm$ 4.74}  \\ \cline{2-4} 
                                                     & \alghota                                       & \multicolumn{1}{c|}{N/A} & \multicolumn{1}{c|}{89.37 $\pm$ 6.97} \\ \hline
\multirow{3}{*}{$\gamma = 0.5$}                                 & {\cellcolor[gray]{.9}{\bf \algp}}                                     & \multicolumn{1}{c|}{\cellcolor[gray]{.9} {\bf 71.89}} & \multicolumn{1}{c|}{\cellcolor[gray]{.9} {\bf 88.89 $\pm$ 3.33}} \\ \cline{2-4} 
                                                     & \alg                                           & \multicolumn{1}{c|}{68.77} & \multicolumn{1}{c|}{80.84 $\pm$ 5.56} \\ \cline{2-4} 
                                                     & \alghota                                       & \multicolumn{1}{c|}{N/A} & \multicolumn{1}{c|}{69.77 $\pm$ 3.89} \\ \hline
\end{tabular}
}
\vspace{-0.1in}
\end{table}

Fig.~\ref{fig:result1} shows \algp's superior test accuracy over benchmarks in personalized tasks with $\gamma=0.5$, highlighting the advantage of personal RIS. This confirms our joint design's efficacy and suggests that smaller, personal RISs outperform a single large RIS, especially in diverse device and multi-task FL scenarios.
Table~\ref{table:test_accuracy} presents results of different non-i.i.d. cases, with lower $\gamma$ indicating more unbalanced data.
Consequently, the performance of the global model drops at $\gamma=0.1$ compared to $\gamma=0.5$.
Notably, \alghota~has the lowest accuracy when $\gamma=0.5$, yet it surpasses \algns~when $\gamma=0.1$. This indicates that the design for personalized FL is more effective when data diversity is greater, even without RIS.
\algp~improves global learning and excels in personalized tasks.
Furthermore, the performance difference between the personalized and global models grows as the non-i.i.d. level increases, demonstrating that \algp~is better suited for handling highly heterogeneity.

\vspace{-0.1in}
\section{Conclusion} 
\label{sec: conclusion}
\vspace{-0.1in}
We have introduced the first personalized OTA-FL using a bi-level optimization multi-task framework with personal RIS assistance.
Our alternating, cross-layer method optimally utilizes communication and computation resources for both global and personalized tasks. 
The proposed algorithm, \algp, can handle non-convex objectives and device heterogeneity, adapting power control, local steps, and RIS settings. 
It outperforms the state-of-the-art algorithms under imperfect CSI scenarios.


\clearpage

\bibliographystyle{IEEEbib}
\bibliography{refs}

\end{document}

%% file: symbols_commands.tex
\newcommand{\Cb}{\mathbb{C}}

\newcommand{\g}{\mathbf{g}}
\newcommand{\Hb}{\mathbf{H}}
\newcommand{\h}{\mathbf{h}}
\newcommand{\hh}{\widehat{h}}

\newcommand{\Rb}{\mathbb{R}}

\newcommand{\sigt}{\widetilde{\sigma}}

\newcommand{\w}{\mathbf{w}}

\newcommand{\x}{\mathbf{x}}

\newcommand{\y}{\mathbf{y}}
\newcommand{\z}{\mathbf{z}}

\newcommand{\0}{\mathbf{0}}

\newcommand{\mc}[1]{\mathcal{#1}}
\newcommand{\mb}[1]{\mathbb{#1}}
\newcommand{\mf}[1]{\mathbf{#1}}

\newtheorem{assum}{Assumption}

\newcommand{\The}{\mathbf{\Theta}}
\newcommand{\theb}{\boldsymbol{\theta}}
\newcommand{\phib}{\boldsymbol{\phi}}

